\documentclass[prd,floatfix,nofootinbib,preprintnumbers]{revtex4}
\usepackage{epsfig,subfigure}
\usepackage{graphicx}

\usepackage{amssymb}

\def\spose#1{\hbox to 0pt{#1\hss}}
\def\lsim{\mathrel{\spose{\lower 3pt\hbox{$\mathchar"218$}}
 \raise 2.0pt\hbox{$\mathchar"13C$}}}
\def\gsim{\mathrel{\spose{\lower 3pt\hbox{$\mathchar"218$}}
 \raise 2.0pt\hbox{$\mathchar"13E$}}}
 \def\be{\begin{eqnarray}}
 \def\ee{\end{eqnarray}}

\newcommand{\nnb}{\nonumber}

\def\lbar{\overline}

\begin{document}
\vskip0.5pc
\parskip 2mm
\parindent 8mm
\indent

\title{Search for new physics from $B\to\pi \phi$}
\vspace{0.5cm}
\author{Jian-Feng Cheng$^a$, Yuan-Ning Gao$^a$,  Chao-Shang Huang$^b$, and Xiao-Hong Wu$^{c}$}
\affiliation{
 $^a$ Center for High Energy Physics, Tsinghua University,
             Beijing 100084,  China\\
 $^b$ Institute of Theoretical Physics, Academia Sinica, P. O. Box 2735,
             Beijing 100080,  China\\
 $^c$  School of Physics,  Korea Institute for Advanced Study, Seoul 130-722, Korea }

\vspace{0.3cm}
\begin{abstract}
\vspace{0.2cm}\noindent We investigate the pure penguin process
$B^-\to \pi^-\phi$ using QCD factorization approach to calculate
hadronic matrix elements to the $\alpha_s$ order in some
well-known NP models. It is shown that the NP contributions in
R-parity conserved SUSY models and 2HDMs are not enough to
saturate the experimental upper bounds for $B\to \phi \pi$. We
have shown that the flavor changing $Z^\prime$ models can make the
branching ratios of $B\to \phi \pi$ to saturate the bound under
all relevant experimental constraints.
\end{abstract}
\maketitle


The process $B\rightarrow \phi \pi$, one of charmless two-body
nonleptonic decays of B mesons, is interesting because it is a
pure penguin process. In particular, there are no annihilation
diagram  contributions for which results obtained with different
methods are quite different~\cite{li,bbns}. Therefore the
calculations of the hadronic matrix elements relevant to the process
are relatively reliable because of no contributions coming from
diagrams of annihilation topology. It proceeds through $b\to
d\bar{s} s$ at the quark level, which is a $b\rightarrow d$ flavor
changing neutral current (FCNC) process. It is sensitive to new
physics (NP) because all contributions arise from the penguin
diagrams in the standard model (SM).

The BaBar collaboration has recently reported the results of
search for $B\rightarrow \phi \pi$~\cite{data}: \be
Br(B^0\rightarrow \phi \pi^0) & = &(0.12\pm 0.13)\times 10^{-6},\\
\nnb & < & 0.28\times 10^{-6},\\ \nnb Br(B^+\rightarrow \phi
\pi^+)& < & 0.24 \times 10^{-6}, \ee which enhance the precision
of measurements but still are of the same order of magnitude,
i.e., $O(10^{-7})$, comparing with the previous
results~\cite{datao} \be
Br(B^0\rightarrow \phi \pi^0) & = &(0.2^{+0.4}_{-0.3}\pm 0.1)\times 10^{-6},\\
\nnb & < & (1.2\pm 0.8)\times 10^{-6},\\ \nnb Br(B^+ \rightarrow
\phi \pi^+)& < & 0.41 \times 10^{-6}. \ee

The SM predictions for these decays modes in both BBNS (QCDF) and
Li et al. (PQCD) methods have been given~\cite{ch} and the results
given in the literature are much smaller than the data although
there are significant disagreements among the literature. The
twist-3 contributions to the decays are predicted to be small and
$Br(B^+\rightarrow \phi \pi^+)=(2-6)\times 10^{-9}$ is given by
using the QCDF method improved in calculating integrals which
contain endpoint singularities, including the twist-3
contributions~\cite{ch}. The agreed results are obtained in QCDF
without including the twist-3 contributions in ref.\cite{gm}. The
theoretical uncertainty in treating integrals which contain
endpoint singularities is roughly $30\%$. Therefore, there is a
room for new physics (NP).

The theoretical study for these decay modes in models beyond the
standard model has also been done in a number of
papers~\cite{ch,gm, bey}. The branching ratio for
$Br(B^+\rightarrow \phi \pi^+)$ has been calculated in constrained
minimal supersymmetric standard model (CMSSM) without imposing the
constraint from $B_s\rightarrow \mu^+\mu^-$~\cite{ch}. In
ref.\cite{gm} calculations are done in MSSM, the topcolor assisted
technicolor model (TC2), and the model with an extra vector like
down quark (VLDQ) respectively. Only for VLDQ $Br(B^0\rightarrow
\phi \pi^0)\sim 10^{-7}$ can be obtained. The analysis in MSSM
uses the values of the mass insertion parameters which are taken
from the paper in 1996~\cite{ggms} and does not consider the
contributions from neutral Higgs boson induced operators. In
ref.\cite{bey} the upper bound of $Br(B^+\rightarrow \phi \pi^+)$
in 2002~\cite{02} is used to constrain parameters in R-parity
violating supersymmetric models.

In the letter we show that $O(10^{-7})$ branching ratio for
$Br(B\rightarrow \phi \pi)$ can not be reached in MSSM and two
Higgs doublet models (2HDM) I, II, and III when all relevant
constraints from experiments are imposed. It can be reached in a
flavor changing $Z^\prime$ model under all relevant experimental
constriants.

To begin with, we record the SM analysis in ref.\cite{ch} briefly.
The $\Delta B = 1$  effective weak Hamiltonian in SM is given by

\begin{equation}\label{Heff}
   {\cal H}_{\rm eff}^{SM} = \frac{G_F}{\sqrt2} \sum_{p=u,c} \!
   \lambda_p \bigg( C_1\,Q_1^p + C_2\,Q_2^p
   + \!\sum_{i=3,\dots, 10}\! C_i\,Q_i + C_{7\gamma}\,Q_{7\gamma}
   + C_{8g}\,Q_{8g} \bigg) + \mbox{h.c.} \,,
\end{equation}
where $\lambda_p=V_{pb}V_{pd}^{*}$, $Q_{1,2}^p$ are the
left-handed current--current operators arising from $W$-boson
exchange, $Q_{3,\dots, 6}$ and $Q_{7,\dots, 10}$ are QCD and
electroweak penguin operators, and $Q_{7\gamma}$ and $Q_{8g}$ are
the electromagnetic and chromomagnetic dipole operators,
respectively. Their explicit expressions can be found in, e.g.,
Ref.~\cite{bbns1}. Due to the flavor and color structures of the
final state $\phi \pi$, the chromomagnetic dipole operator
$Q_{8g}$ does not contribute to the decays, and the tree operators
$Q_{1,2}$ contribute only through electromagnetic corrections
which is small numerically. In QCDF up to the order of $\alpha_s$,
the decay amplitude for $B^-\to \pi^-\phi$ is~\cite{ch}
\begin{eqnarray}\label{amp}
&&A(B^-\to \pi^-\phi)\nonumber\\
&=&\sqrt{2}\,A(B^0\to \pi^0\phi)\nonumber \\
&=& {G_F\over\sqrt{2}}\sum\limits_{p=u,c}\lambda_p \left[ a_3+a_5
-{1\over 2 }\left(a_7^p+a_9^p\right)\right]\, f_\phi m_\phi
F^{B\to\pi}_+(m^2_\phi)~2~\epsilon^\phi_L\cdot p_B ,
\end{eqnarray}
where \begin{eqnarray} a_3(\pi\phi) &=& C_3+\frac{1}{N}C_4 +
\frac{\alpha_s}{4\pi}\frac{C_F}{N}\,C_4\,F,
\label{a3}\\
a_5(\pi\phi) &=& C_5+\frac{1}{N}C_6 +
\frac{\alpha_s}{4\pi}\frac{C_F}{N}\,C_6\,(-F-12),
\label{a5}\\
a^p_7(\pi\phi)&=& C_7+{C_8\over N}+{\alpha_s\over
4\pi}\frac{C_F}{N}\,C_8\,(-F-12)+ {\alpha_{\rm em}\over 9 \pi}
P^p_{\rm em} (C_1+3 C_2), \label{a7} \\
a^p_9(\pi\phi) &=& C_9+\frac{1}{N}C_{10} +
\frac{\alpha_s}{4\pi}\frac{C_F}{N}\,C_{10}\,F + {\alpha_{\rm em
}\over 9 \pi }P^p_{\rm em} (C_1+3 C_2)\label{a9}\,,
\end{eqnarray}
Due to the almost completely cancellations of the two terms in the
Wilson coefficient combinations, $C_3+C_4/N_c$ and $C_5+C_6/N_c$,
$a_9$ is the biggest among $a_i$'s, i.e., the contributions of the
electroweak penguin operators dominate. With values of input
parameters in ref.\cite{ch}, $Br(B^+\rightarrow \phi \pi^+)=4.5
\times 10^{-9}$ is obtained.

We now turn to MSSM and 2HDM. The effective Hamiltonian in 2HDM
and MSSM can be written as~\cite{ch,chw}
\begin{eqnarray}
\label{newh} {\cal H}_{\rm eff} &=& {\cal H}_{\rm eff}^{SM}+{\cal
H}_{\rm eff}^{new}, \\ \nnb
 {\cal H}_{\rm eff}^{new} &=& \frac{G_F}{\sqrt2} \sum_{p=u,c} \!
   V_{pb} V^*_{pd} \bigg(\!\sum_{i=11,\dots, 16}\![ C_i\,Q_i+ C_i^\prime\,Q_i^\prime]
   \nonumber \\&& + \!\sum_{i=3,\dots, 10}\!C_i^\prime\,Q_i^\prime
   + C_{7\gamma}^\prime\,Q_{7\gamma}^\prime
   + C_{8g}^\prime \,Q_{8g}^\prime \, \bigg) + \mbox{h.c.}, \,
\end{eqnarray}
where $Q_{11}$ to $Q_{16}$ are the neutral Higgs penguin operators
and their explicit forms can be found in Ref.~\cite{ch,chw} with
the substitution $s\to d$. The primed operators, the counterpart
of the unprimed operators, are obtained by replacing the chirality
in the corresponding unprimed operators with opposite ones. In
ref.\cite{gm} the neutral Higgs penguin operators $Q_i, i=11, ...,
16$ are not considered.

 The contributions of new operators to the
decay amplitude for $B^-\to \pi^-\phi$ is
\begin{eqnarray}\label{ampn}
&&A^{new}(B^-\to \pi^-\phi)\nonumber\\
&=&\sqrt{2}\,A^{new}(B^0\to \pi^0\phi)\nonumber \\
&=& {G_F\over\sqrt{2}}\sum\limits_{p=u,c}\lambda_p \left[
a_3^\prime+a_5^\prime -{1\over 2 }\left(a_7^{\prime p}+a_9^{\prime
p}\right)+ r_\chi^\phi\,(a_{11}+a_{13})\right]\, f_\phi m_\phi
F^{B\to\pi}_+(m^2_\phi)~2~\epsilon^\phi_L\cdot p_B ,
\end{eqnarray}
where we have neglected $O(\frac{m_\phi^2}{m_B^2})$ terms,
\begin{equation}\label{rKdef}
   r_\chi^\phi(\mu)
   =  {m_B\over 4 \epsilon \cdot p_B}\, {f_\phi^T \over f_\phi}
   \,,
\end{equation}
$a_i^\prime, i=3,5,7,9,$ come from the contributions of the primed
operators and their expressions can be obtained by replacing the
Wilson coefficients in the corresponding unprimed operators with
primed ones, $a_{11,13}$ come from the contributions of neutral
Higgs boson induced operators which we shall discuss later on.

In MSSM new contributions to $a_i, i=3,5$ and the contributions to
$a_i^\prime, i=3,5$ come from the gluino-sbottom loop and depend
on the mass insertion parameters $(\delta^d_{AA})_{13}, A=L,R$.
The constraints on $(\delta^d_{AA})_{13}, A=L,R$ from the mass
difference $\Delta M_d$ have been reanalyzed, using NLO QCD
corrections of Wilson coefficients and recent lattice calculations
of the B parameters for hadronic matrix elements of local
operators~\cite{beci}. Depending on the average mass of squarks
and the gluino mass as well the CKM angle $\gamma$,
$(\delta^d_{AA})_{13}, A=L,R$, is constrained to be $O(10^{-2})$
in the $(\delta^d_{LL})_{13}=(\delta^d_{RR})_{13}$ case which is
assumed in ref.\cite{gm}. So in this case the new contributions to
$a_i, i=3,5$ and the contributions to $a_i^\prime, i=3,5$ are
negligibly small, compared with the SM. Only for the case of
single $(\delta^d_{LL})_{13}$ (or $(\delta^d_{RR})_{13}$) non
zero, $(\delta^d_{LL})_{13}$ (or $(\delta^d_{RR})_{13}$) can reach
$O(10^{-1})$ under the constraint from the mass difference $\Delta
M_d$~\cite{beci} and consequently the SUSY contributions can make
the branching ratios of $B\to \phi\pi$ reached $O(10^{-8})$.

The contributions of neutral Higgs boson induced operators,
$a_{11,13}$, are as follows.
\begin{eqnarray}
&&a_{11,13} = -{\alpha_s\over 4\pi } \, {C_F\over N}\, (f^I_s +
f^{II}_s ) \, C_{Q_{12,14}}\,, \\ \nnb
\end{eqnarray}
where
\begin{eqnarray}
&&f^I_s = - 2\,\int ^1 _0 du  \big[ \ln^2 u + 2\ln u - 2{\rm Li}_2
(u) \big]
\, \phi_s(u)\nonumber\\
&&\hspace{0.4cm} + 2 \int^1 _0 du \int^{m_b}_0 dk \int db \, \ln
\left[ \sqrt{ {4k^2_T\over m_b^2 } +u^2 } +u \right] \, J_0 ( bk )
{\cal P}_s ( u,b) \\
&& f^{II}_s =   { 2\pi\,m_B \, f_\pi f_B\over F^{B\to
\pi}_+{m^2_\phi}} \int [du][db]\,
\delta ^2 (b_1 +b_2 )\, b\, {\cal P}_B(\xi,b)\, {\cal P}_s(v,b_2)\nonumber\\
&&\hspace{0.4cm}\times \big[ \mu_p\,
(u+v)\,{\cal P}_p(u,b_1)+m_B\,(\xi-v) \,{\cal P}(u,b_1)\big] \nonumber\\
&&\hspace{0.4cm} \times \left[ \theta(b_2-b)I_0(b\sqrt{u\xi}\,m_
B)\,K_0(b\sqrt{u\xi} \, m_B) +\theta(b-b_2 )\, I_0
(b\sqrt{u\xi}\,m_B )\,
K_0 (\sqrt{-uv}m_b b_2 ) \right]\,. \nonumber\\
\ \
\end{eqnarray}
Note that there are no contributions of neutral Higgs boson
induced operators at the leading order in the $\alpha_s$ expansion
in the approximation omitting $O(m_{\phi}^2/m_{B}^2)$ terms because of their Dirac structure.


In MSSM and 2HDM model I, II, $C_{11,13}^{(\prime)}(m_W)\sim
0.037$ because of the constraint from $B_d\rightarrow
\mu^+\mu^-$~\cite{cghw}. So the $\alpha_s$ corrections from
neutral Higgs boson induced operators are negligible.


In model III 2HDM there are neutral Higgs boson-mediated FCNC at
the tree level~\cite{m3,cs,at}. The Yukawa Lagrangian for quarks
can be written as~\cite{cck}
\begin{eqnarray}
{\cal L}_Y &=& -\bar{U} M_U U - \bar{D} M_D D
   - \frac{g}{2M_W} (H^0\cos\alpha - h^0 \sin\alpha)
     \bigg (\bar{U} M_U U + \bar{D} M_D D \bigg ) \nonumber \\
&+& \frac{ig}{2 M_W} G^0 \left(\bar{U} M_U \gamma^5 U
                             - \bar{D} M_D \gamma^5 D \right)\nonumber \\
&+& \frac{g}{\sqrt{2}M_W} G^- \bar{D} V^\dagger_{\rm CKM} \bigg [
     M_U \frac{1}{2}(1+\gamma^5) - M_D \frac{1}{2}(1-\gamma^5) \bigg]
 U \nonumber \\
&-& \frac{g}{\sqrt{2}M_W} G^+ \bar{U} V_{\rm CKM} \bigg [
     M_D \frac{1}{2}(1+\gamma^5) - M_U \frac{1}{2}(1-\gamma^5) \bigg]
 D  \nonumber \\
&-& \frac{H^0 \sin\alpha + h^0\cos\alpha }{\sqrt{2}} \bigg[
  \bar{U} \bigg( {\hat\xi}^U \frac{1}{2}(1+\gamma^5) + {\xi}^{U\dagger}
 \frac{1}{2}(1-\gamma^5)
         \bigg ) U  \nonumber \\
&& +\bar{D} \bigg( {\xi}^D \frac{1}{2}(1+\gamma^5) +
 {\xi}^{D\dagger} \frac{1}{2}(1-\gamma^5)
         \bigg ) D
  \bigg ]  \nonumber \\
&+& \frac{i A^0}{\sqrt{2}} \bigg [
    \bar{U} \bigg( {\xi}^U
           \frac{1}{2}(1+\gamma^5) -{\xi}^{U\dagger}
           \frac{1}{2}(1-\gamma^5)
        \bigg ) U
   -\bar{D} \bigg( {\xi}^D
           \frac{1}{2}(1+\gamma^5) - {\xi}^{D\dagger}
           \frac{1}{2}(1-\gamma^5)
        \bigg ) D
     \bigg ] \nonumber \\
&-&  H^+ \bar{U} \bigg[ V_{\rm CKM} { \xi}^D
             \frac{1}{2}(1+\gamma^5) -
   {\xi}^{U\dagger} V_{\rm CKM}
             \frac{1}{2}(1-\gamma^5) \bigg] D
  \nonumber \\
&-&  H^- \bar{D} \bigg[ {\xi}^{D\dagger} V^\dagger_{\rm CKM}
   \frac{1}{2}(1-\gamma^5) -
      V^\dagger_{\rm CKM} {\xi}^U \frac{1}{2}(1+\gamma^5) \bigg] U
 \;\;,
\label{rule}
\end{eqnarray}
where $U$ and $D$ represents the mass eigenstates of $u,c,t$
quarks and $d,s,b$ quarks respectively, the matrices $\xi^{D,U}$
are in general non-diagonal which parameterize the couplings of
Higgs to quarks. The Yukawa Lagrangian for leptons can similarly
be written.

With the above Lagrangian, the Wilson coefficients of $Q_i,
i=11,...,16,$ relevant to $b \to d\bar{s} s$ are easily obtained:
\be
C_{11,13}(m_W)&=&\frac{\sqrt{2}}{G_F(-\lambda_t)}(-\frac{1}{8}\xi_{bd}\xi_{ss})
(\frac{\sin^2\alpha}{m_H^2}+\frac{\cos^2\alpha}{m_h^2}\mp\frac{1}{m_A^2})\frac{m_b}{m_s}\,,\nnb \\
C_{12,14,15,16}(m_W)&=&0\,,\nnb\\
C^\prime_{11,13}&=&C_{11,13}\,,~~~~~~C^\prime_{12,14,15,16}(m_W)=0\,,
\ee where we have assumed $\xi_{ij}$ is symmetric and real for
simplicity. The parameters $\xi_{bd,ss}$ of the model can be
estimated from experiments. We can extract $\xi_{bd}$ from the
mass difference
$\Delta M_{B_d}$ of neutral $B_d$ mesons. 
 In order to determine $\xi_{ss}$ we use the data of $\tau\to \mu
\mu^+\mu^-,\, B_d\to \mu^+\mu^-$ and $\tau\to \mu P(P=\pi^0, \eta,
\eta^\prime)$. The parameter $\xi_{bd}$ has been estimated from
the measured $B_d - \bar{B}_d$ mass difference, and $\xi_{db} \leq
7.3\times 10^{-6}$ 
is given~\cite{diaz04}. We reanalyze the mass difference and get
$\xi_{db} \leq 0.8 \times 10^{-4}$ which is much larger than that
in~\cite{diaz04}.
 With the latest limits from experiments,
$Br(B_d \to \mu \mu) \leq 3.9 \times 10^{-8}$~\cite{bdmumucdf} and
$Br(\tau \to \mu \mu \mu) \leq 1.9 \times
10^{-7}$~\cite{tau3mubabar} at 90\% C.L., we obtain the updated
results of the upper bounds of $\xi_{\mu\mu}$ and $\xi_{\mu\tau}$,
$\xi_{\mu\mu} \leq 0.34$ and $\xi_{\mu\tau} \leq 0.004$. Finally
we can determine the bound on $\xi_{ss}$ from $\tau \to \mu \pi^0
(\eta, \eta^\prime)$ decays~\cite{wjli05} and the result is
$\xi_{ss} \leq 0.26$. Using the bounds of $\xi_{db}$ and
$\xi_{ss}$, we deduce that $C_{13}(m_W)$ can roughly reach $0.1$
with $m_A = m_H = 300 GeV, \; m_h=120 GeV$,
which leads to that the $\alpha_s$ corrections from neutral Higgs
boson induced operators can contribute to the branching ratio of
$Br(B^+\rightarrow \phi \pi^+)$ by about $2.0\times 10^{-10}$
which is much smaller than that in SM and consequently negligible.


Finally we consider flavor changing $Z^\prime$ Models. Because the
contributions of electroweak operators dominate in SM it is
expected that the new contributions from flavor changing
$Z^\prime$ models in which $Z^\prime$ mediates vector and axil
vector interactions would enhance the branching ratios of $B\to
\phi \pi$ significantly. The flavor changing $Z^\prime$ Models
have been extensively studied~\cite{zmod}. For our purpose we
write the $Z^\prime$ interaction Lagrangian in the gauge basis as
\begin{eqnarray} \label{lagran}
{\cal L}^{Z'} &=&  - g^\prime J'_{\mu} Z'^{\mu} ~, \\
J'_{\mu} &=&  \sum_{i,j} {\lbar \psi_i^I} \gamma_{\mu}
  \left[ (\epsilon_{\psi_L})_{ij} P_L + (\epsilon_{\psi_R})_{ij} P_R \right]
    \psi^I_j ~,
     \end{eqnarray}
where $g^\prime$ is the gauge coupling constant of the
$U(1)^\prime$ group at the $M_W$ scale\footnote{We neglect the
renormalization running effects from the $M_Z^\prime$ scale to the
$M_W$ scale through the paper due to uncertainties in the
parameters of the models.}, the sum extends over all fermion
fields in the SM. There are in general FCNCs at the tree level in
the Lagrangian (\ref{lagran}). In the mass eigenstate basis, the
chiral $Z^\prime$ coupling matrices are respectively given by \be
B^X_u &\equiv& V_{u_X} \epsilon_{u_X} V_{u_X}^{\dagger} ~, ~~
B^X_d \equiv V_{d_X} \epsilon_{d_X} V_{d_X}^{\dagger} ~, ~~ (X =
L,R) \ee where $V_{u_X}$ and $V_{d_X}$ are the transformation
matrices which make up-type quarks and down-type quarks into the
mass eigenstates from the gauge eigenstates respectively. The
usual CKM matrix is given by $V_{CKM}=V_{u_L}V_{d_L}^\dagger$. We
assume that the first two generation diagonal elements of $B^X_q
(q=u,\,,d)$ are equal: $B_{dd}^X=B^X_{ss}$~\cite{cdj,lang}. The
effective Hamiltonian of the $b\to d q\bar{q}$ transitions due to
the $Z^\prime$ mediation is \be H_{\rm eff}^{Z'} &=&\frac{2
G_F}{\sqrt{2}} \left(\frac{g^\prime M_Z}{g_Z M_{Z'}}\right)^2
B^{L*}_{db}
   ({\bar b}d)_{V-A} \sum_q \left( B^L_{qq} ({\bar q}q)_{V-A}
 + B^R_{qq} ({\bar q}q)_{V+A} \right) + \mbox{h.c.} ~,
\label{eqn:Heff1} \ee where $M_{Z^\prime}$ is the mass of
$Z^\prime$ which is larger than $850 GeV$ for $g^\prime\sim
g_Z$~\cite{gri}, \be g_Z=\frac{e}{\sin \theta_W \cos\theta_W}\,.
\ee Thus, the $Z^\prime$ mediated FCNC interaction would induce
the color singlet QCD and electroweak penguin operators. It is
straightforward to get the contributions to the Wilson
coefficients of those operators from above effective Hamiltonian,
(\ref{eqn:Heff1}). Then the effective Hamiltonian relevant to
$B\to \phi \pi$ in a $Z^\prime$ model can be written as
\begin{eqnarray}
\label{newh} {\cal H}_{\rm eff} &=& {\cal H}_{\rm eff}^{SM}+{\cal
H}_{\rm eff}^{Z^\prime}, \\ \nnb
 {\cal H}_{\rm eff}^{Z^\prime} &=& - \frac{G_F}{\sqrt2}
   \lambda_t \bigg(
   \!\sum_{i=3,5,7,9}\! C_i^{Z^\prime}\,Q_i  \bigg) + \mbox{h.c.} \,, \ee
where $\lambda_t=V_{tb}^* V_{td}$, and \begin{eqnarray} &&
C^{Z\prime}_{3(5)} = - \frac{2}{3 \lambda_t} \left(\frac{g_2
M_Z}{g_1
    M_{Z'}}\right)^2 B^{L*}_{db} \left(B^{L(R)}_{uu} + 2
B^{L(R)}_{dd} \right) \nonumber\\
&& C^{Z^\prime}_{9(7)} = -\frac{4}{3 \lambda_t} \left(\frac{g_2
M_Z}{g_1
    M_{Z'}}\right)^2 B^{L*}_{db} \left(B^{L(R)}_{uu} -
B^{L(R)}_{dd} \right).
\end{eqnarray}
The chiral $Z^\prime$ couplings are subjected to constraints from
relevant experimental measurements. The constraint to $B_{bd}^L$ has
been analyzed~\cite{cdj,hv}.  It is shown~\cite{cdj} that
the observed $\Delta M_{B_d}$ and $\sin 2\beta$ leads to
\begin{eqnarray}\label{cn1}
 y|{\rm Re}(B_{db}^{L})^2| < 5 \times 10^{-8}\,,
\end{eqnarray}
where $y=\left(\frac{g^\prime M_Z}{g_Z M_{Z^\prime}}\right)^2$, if
only left-handed couplings are considered. Taking $y\sim
10^{-3}$~\cite{cdj}, one has \be B_{db}^{L}\sim 0.7\times
10^{-2}\label{db} \ee.

 When both left-handed and right handed couplings are included,
the constraints are
\begin{eqnarray}
&&  y| {\rm Re}[(B_{db}^{L})^2 + (B_{db}^{R})^2]
- 3.8 {\rm Re} ( B_{db}^{L}B_{db}^{R})| <  5 \times 10^{-8}\,, \\
&&  y| {\rm Im}[(B_{db}^{L})^2 + (B_{db}^{R})^2] - 3.8 {\rm Im} (
B_{db}^{L}B_{db}^{R})| <  5 \times 10^{-8}\,,
\end{eqnarray}
which are not as stringent as eq. (\ref{cn1}) because of the
possible cancellation among different terms.

Using the latest observed $\Delta M_{B_s}$,
ref.\cite{hv} shows that
\begin{eqnarray}
  B^{L*}_{sb}
\sim& 10^{-2} \label{sb}
\end{eqnarray}

To solve the $\pi K$ puzzle some people have examined new physics in
 the electrweak penguin sector, in particular, the $Z^\prime$
 models~\cite{bur,lang}. Using the experimental data of $Br(B\to \pi
 K)$ in and before 2003, one has~\cite{lang} \be \label{rcrn}
 R_c=1.15\pm0.12,~~~~R_n=0.78\pm 0.10,\ee which should be roughly equal to one
 in the SM. It is shown that to explain the deviations a constraint on
$B^{L*}_{sb} B^X_{dd}$ must be imposed~\cite{lang}
\be \label{sbdd} \xi^{LL}\equiv y |\frac{B^{L*}_{sb}
 B^X_{dd}}{V_{tb}^*V_{ts}}|\approx
0.01 \,(solution\; A)\, or\, 0.019\,(solution\; B) \ee in the case of
 $\xi^{LL}=\xi^{LR}$. We reanalyze
the constraint using the new data of $Br(B\to \pi K)$ in this year~\cite{data}. Accoding to
 the new data, one has~\cite{wuz}, instead of eq.(\ref{rcrn}), \be
 R_c=1.11\pm 0.08,~~~~R_n=1.00\pm 0.07\,.\ee With the values, the
 solution B in ref.\cite{lang} is excluded in $1\sigma$ bounds, but
 the solution A remains survived. So from eqs.(\ref{sb},\ref{sbdd}), we have
\begin{eqnarray}\label{dd}
B^X_{dd} &\sim & 10\, .
\end{eqnarray}

The $Z^\prime$ flavor changing couplings also contribute to
$B^0\to \pi^0\pi^0$. The constraint from $Br(B^0\to \pi^0\pi^0)$
is:
\begin{eqnarray}
0.129 &<& \left| 0.0887+0.0820i + \left( 1.277 B^L_{dd}
-B^R_{dd}\right) B^{L*}_{db} \right|^2 < 0.178  \ee Form eqs.
(\ref{db},\ref{dd}), the above constraints are satisfied.

Now we obtain the branching ratio of $B\to \pi \phi$ in the
$Z^\prime$ models for the values of parameters which satisfy the
constraints discussed above: \be
 Br(\pi\phi) &=& 1.17\times 10^{-5} \left| 0.0184 + 0.00607i -
\left( B^L_{dd} +B^R_{dd}\right) B^{L*}_{db} \right|^2 \sim
10^{-7},
\end{eqnarray}
which is in agreement with the recent data.

In summary, we have studied the pure penguin process $B^-\to
\pi^-\phi$ using QCD factorization approach to calculate hadronic
matrix elements to the $\alpha_s$ order in some well-known NP
models. We have shown that the NP contributions in R-parity
conserved SUSY models and 2HDMs are not enough to saturate the
experimental upper bounds for $B\to \phi \pi$. We have also shown
that the flavor changing $Z^\prime$ models can make the branching
ratios of $B\to \phi \pi$ to saturate the bounds under all relevant
experimental constraints. Therefore, if the data will remain in
$O(10^{-7})$ in the future it will give a signal of NP effects and
provide a clue to discriminate well-known NP models.

{\bf \Large Acknowledgement}

One of the authors (C.-S. Huang) would like to thank the Korea Institute
for Advanced Study (KIAS) for warm hospitality during his visit.
The work was supported in part by the Natural Science Foundation
of China (NSFC), grant 10435040, grant 90503002, and grant
10225522.

\end{document}